\documentclass[journal]{IEEEtran}

\ifCLASSINFOpdf
\else
\fi
\usepackage{xcolor}
\usepackage{graphicx}
\usepackage{placeins}
\usepackage{amssymb}% http://ctan.org/pkg/amssymb
\usepackage{pifont}% http://ctan.org/pkg/pifont
\newcommand{\xmark}{\ding{55}}%
\usepackage{amsmath}
\usepackage{multirow}
\usepackage[ruled,linesnumbered,noresetcount]{algorithm2e}
\begin{document}
\title{Feature Learning and Ensemble Pre-Tasks Based Self-Supervised Speech Denoising and Dereverberation}

\author{Yi~Li,~\IEEEmembership{Student Member,~IEEE,}
        ShuangLin~Li,~\IEEEmembership{Student Member,~IEEE,}\\
        Yang~Sun,~\IEEEmembership{Member,~IEEE,}
        and~Syed~Mohsen~Naqvi,~\IEEEmembership{Senior~Member,~IEEE}% <-this % stops a space
\thanks{Yi Li, ShuangLin Li, and Syed Mohsen Naqvi are with the Intelligent Sensing and Communications Group, School of Engineering, Newcastle University, Newcastle upon Tyne NE1 7RU, U.K. (e-mails: y.li140, l.shuanglin2, mohsen.naqvi@newcastle.ac.uk)
}% <-this % stops a space
\thanks{Yang Sun is working with the Big Data Institute, University of Oxford, Oxford OX3 7LF, U.K. (e-mail: Yang.sun@bdi.ox.ac.uk)}% <-this % stops a space

\thanks{E-mail for correspondence: y.li140@newcastle.ac.uk}
}
\markboth{Journal of \LaTeX\ Class Files,~Vol.~14, No.~8, August~2015}%
{Shell \MakeLowercase{\textit{et al.}}: Bare Demo of IEEEtran.cls for IEEE Journals}

\maketitle

% in the abstract or keywords.Besides, in the vanilla DNN-based methods, the temporal information cannot be fully utilized.
\begin{abstract}

Self-supervised learning (SSL) achieves great success in monaural speech enhancement, while the accuracy of the target speech estimation, particularly for unseen speakers, remains inadequate with existing pre-tasks. As speech signal contains multi-faceted information including speaker identity, paralinguistics, and spoken content, the latent representation for speech enhancement becomes a tough task. In this paper, we study the effectiveness of each feature which is commonly used in speech enhancement and exploit the feature combination in the SSL case. Besides, we propose an ensemble training strategy. The latent representation of the clean speech signal is learned, meanwhile, the dereverberated mask and the estimated ratio mask are exploited to denoise and dereverberate the mixture. The latent representation learning and the masks estimation are considered as two pre-tasks in the training stage. In addition, to study the effectiveness between the pre-tasks, we compare different training routines to train the model and further refine the performance. The NOISEX and DAPS corpora are used to evaluate the efficacy of the proposed method, which also outperforms the state-of-the-art methods.

\end{abstract}


% \fi
\IEEEpeerreviewmaketitle

\section{INTRODUCTION}
\IEEEPARstart{S}{peech} signals recorded in an enclosure with a single and distant microphone are subject to reverberation, which degrade the speech intelligibility in audio signal processing algorithms \cite{tomo}. Thus, monaural speech enhancement comprising denoising and dereverberation is the task of providing the enhanced speech signal and improving the speech quality. Recently, speech enhancement research has seen rapid progress by employing deep learning techniques for several applications such as mobile phones, Voice over Internet Protocol (VoIP), and speech recognition \cite{TAI}. 

Two key challenges in monaural speech enhancement are the gain of clean targets and mismatched training and testing conditions \cite{ssl}. Firstly, contemporary supervised monaural speech enhancement relies on the availability of many paired training examples, which is expensive and time-consuming to produce. This limitation is particularly acute in specialized domains like biomedicine, where crowdsourcing is difficult to apply \cite{ssl4}. Self-supervision has emerged as a promising paradigm to overcome the annotation bottleneck by automatically generating noisy training examples from unlabeled data. In particular, task-specific self-supervision converts prior knowledge into self-supervision templates for label generation, as in distant supervision \cite{ranlp}, data programming \cite{datap}, and joint inference \cite{jointi}. Secondly, the speech enhancement performance is degraded when an acoustic mismatch happens between the training and testing stages. The mismatches could occur when the model is trained on data generated with the unseen speakers, noise types, and SNR levels. In such mismatches, the ability to use the recorded test mixtures in supervised learning (SL) methods to improve the performance in the unseen test configurations is limited. Thus, the recent self-supervised learning (SSL) research is rapidly developed to solve these challenge in supervised speech enhancement.

In recent years, many SSL approaches have been proposed to address the monaural speech enhancement problem. Generally, the technique needs to model the input feature map into meaningful continuous latent representations containing the desired speech information \cite{hubert}. Then, to further improve the speech enhancement performance, the model needs to capture the clean speech information from the learned representation. The clean speech examples used in the pre-training are unseen from the downstream training. Therefore, the ability of the trained model to process the unseen data is improved. One crucial insight motivating this work is the importance of consistency of the targets, not just the correctness, which enables the model to focus on modelling the relationship between the clean speech signal and the noisy mixture. In further research, the well-trained models are evaluated on artificially reverberated datasets to show the dereverberation performance in SSL study \cite{ssl6}. Inspired by our previous work \cite{Two, IET, dsp}, in this paper, an SSL-based method is proposed for speech enhancement problem in real reverberant environments because it is highly practical \cite{ssl}.

% Inspired by \cite{pwgan}\cite{vogan}, we evaluates spectrum in various scales and provide a combination with the best performance.
The contributions of the paper are threefold:

$\bullet$ Two pre-tasks with self-training are proposed to solve the speech enhancement problem. Firstly, we use an autoencoder to learn a latent representation of clean speech signals and autoencoder on noisy mixture with the shared representation of the clean examples. Second, to address the speech enhancement problem with the reverberant environment, the dereverberation mask (DM) and the estimated ratio mask (ERM) are applied in the masking module. The learned latent representation and the masking module are ensemble to estimate the target speech and noisy mixture spectra.

$\bullet$ The latent representation and the masking module share the model but extract different desired information from the feature maps. Therefore, to study the effectiveness between the pre-tasks, we provide different training routines and further use the information obtained from one pre-task to train the other one.

$\bullet$ Various features are individually extracted from the spectra and the performance of each feature is evaluated in the SSL case. Furthermore, to the best of out knowledge, the feature combination is firstly proposed in the SSL-based speech enhancement study to refine the performance. 
% extracted from the clean speech and noisy mixture spectra to train the autoencoders to .

\section{RELATED WORK}
\subsection{Training Targets}
In the reverberant environments, the convolutive mixture is usually generated with the RIRs for reverberant speech and interference $h_{s}(m)$ and $h_{i}(m)$ at discrete time $m$ as:
\begin{equation}
y(m)=s(m) * h_{s}(m)+i(m) * h_{i}(m)
\end{equation}
where ‘$\ast$’ indicates the convolution operator. The desired speech signal, the interference and the reverberant mixture are presented as $s(m)$, $i(m)$, and $y(m)$, respectively. By using the short time Fourier transform (STFT), the mixture is shown as:
\begin{equation}
Y(t, f)=S(t, f) H_{s}(t, f)+I(t, f) H_{i}(t, f)
\end{equation}
where $S(t, f)$, $I(t, f)$ and $Y(t, f)$ denote the STFTs of speech, interference, and mixture at time $t$ and frequency $f$, respectively. Besides, the RIRs for speech and interference are presented as $H_s(t,f)$ and $H_i(t,f)$ respectively. In speech enhancement problem, the aim is to reconstruct the spectrum of the clean speech by using the ideal time-frequency (T-F) mask $M(t, f)$ as:
\begin{equation}
S(t, f)=Y(t, f) M(t, f)
\end{equation}
Generally, the mask $M(t, f)$ is a ratio mask. For example, in our previous work \cite{Two, IET}, the DM and ERM are proposed to estimate the target speech from the reverberant mixture in a two-stage structure. There are two signal approximation (SA)-long short-term memory (LSTM) networks i.e., DM$\_$LSTM and ERM$\_$LSTM which individually trains the DM and ERM. The DM is defined as:
\begin{equation}
DM\left ( t,f \right )= \left [ S\left ( t,f \right )+I\left ( t,f \right ) \right ] Y\left ( t,f \right )^{-1}
\end{equation}
Then, the estimated dereverberated mixture ${\hat{Y}_{d}}\left ( t,f \right )$ is obtained from the output layer of the first network DM$\_$LSTM as:
\begin{equation}
{\hat{Y}_{d}}\left ( t,f \right )=Y\left ( t,f \right ) \widehat{DM}\left ( t,f \right )
\end{equation}
where $\widehat{DM}\left ( t,f \right )$ is the estimated DM. Even though, in practice, obtaining the dereverberated mixtures is very challenging \cite{complexirm}. Therefore, in the second network ERM$\_$LSTM, the ERM is exploited to better model the relationship between the clean speech signal and the estimated dereverberated mixture due to the sequentially trained network structure.
\begin{equation}
\widehat{ERM}\left ( t,f \right )=\frac{{\mid}S(t,f){\mid}}{{\mid}{\hat{Y}_{d}}\left ( t,f \right ){\mid}}.
\end{equation}
The final reconstructed speech signal can be obtained with the estimated $M(t, f)$, i.e., the multiplication of $\widehat{DM}\left ( t,f \right )$ and $\widehat{ERM}\left ( t,f \right )$ as:
\begin{equation}
\hat{S}(t, f)=\widehat{ERM}\left ( t,f \right ) \widehat{DM}\left ( t,f \right )Y\left ( t,f \right )
\end{equation}
However, the two-stage structure suffers a limitation, its computational cost is almost doubled compared to the single-stage model methods. Therefore, in this work, the proposed masking module consists of two T-F masks and is trained as one of pre-tasks in the single-stage model to efficiently improve the speech enhancement performance.
\subsection{Features}
According to \cite{featurezuoyong}, it is well-known that extracted features as input and learning machines play a complementary role in the monaural speech enhancement problem. Therefore, we select five commonly-used features in speech enhancement and provide a brief introduction for them. The complementary feature set of these features has been proved to show stable performance in various test conditions and outperforms each of its components significantly \cite{featureset}.
\subsubsection{Spectrogram}
Recently, the spectrogram has been proved to be a crucial representation for speech enhancement problem with time-frequency decomposition \cite{CSA1}. The spectrogram consists of 2D images representing sequences of short-time Fourier transform (STFT) with time and frequency as axes, and brightness representing the strength of a frequency component at each time frame. In the speech enhancement problem, the noisy mixture spectrogram is fed into the model producing an enhanced speech spectrogram.
\subsubsection{MFCC}
In the mel frequency cepstral coefficients (MFCC) feature extraction, the noisy mixture is passed through a first-order FIR filter in the pre-emphasis stage to boost the high-band formants \cite{mfcc}. As one of the most commonly used features in the speech enhancement problem, the MFCC provides a spectral representation of speech that incorporates some aspects of audition \cite{mfcc1}. Implementation of the spectral feature mapping technique using MFCC features has the advantage of reducing the length of the input feature vector.
\subsubsection{AMS}
Amplitude modulation spectrograms (AMS) are motivated by psycho-physical and psycho-physiological findings on the processing of amplitude modulations in the auditory system of mammals \cite{ams}. Consequently, they have originally been exploited in binaural speech enhancement problem to extract the target speech with spatial separation \cite{ams}. For single-channel speech enhancement with signal-to-noise ratio (SNR) estimation, AMS features are combined with a modulation domain Kalman filter \cite{ams3}. Besides, in reverberant environments, AMS features perform competitive compared to simple spectrogram \cite{ams4}. 
\subsubsection{RASTA-PLP}
In \cite{PLP2}, relative spectral transform and perceptual linear prediction (RASTA-PLP) is first introduced to speech processing. In speech enhancement problem, an overlap-add analysis technique is used to the cubic root of the power spectrum of noisy speech, which has been filtered and then cubed \cite{RASTA}. RASTA-PLP is an extension of perceptual linear prediction (PLP) and the only different from the PLP, is that a band pass filter is added at each sub band \cite{RASTA2}.
% The RASTA processing suppresses the spectral components outside the typical modulation spectrum of speech. However, the noisy phase is preserved for the speech reconstruction . 

\subsubsection{cochleagram}
As a form of spectrogram, the cochleagram assigns a false colour which displays spectra in color recorded in the visible or non-visible parts of spectra to each range of sound frequencies. In speech enhancement problem, the cochleagram exploits a gammatone filter and shows better reveal spectral information than the conventional spectrogram \cite{coch}. The resulting time–frequency feature map provides more frequency components in the lower frequency range with narrow bandwidth and fewer frequency components in the upper frequency range with wide bandwidth, thereby revealing more spectral information than the feature map from the conventional spectrogram \cite{coch}.
\subsection{Self-Supervised Speech Enhancement}
SSL-based speech enhancement involves pre-training a latent representation module on limited clean speech data with an SL objective, followed by large-scale unlabelled data with an SSL objective \cite{ssl}. The latent representation of the clean speech is commonly used as the training target in SSL studies \cite{ssl, ssl5}. The learned representation can capture important underlying structures from the raw input, e.g., intermediate concepts, features, or latent variables that are useful for the downstream task. Following the increasing popularity within the speech enhancement problem, some attempts have been done to extend SSL to discover audio and speech representations \cite{hubert, ssl7}. For example, authors introduce a contrastive learning approach towards self-supervised speech enhancement \cite{cl}. The speaker-discriminative features are extracted from noisy recordings, favoring the need for robust privacy-preserving speech processing. Nevertheless, applying SSL to speech remains particularly challenging. Speech signals, in fact, are not only high-dimensional, long, and variable-length sequences, but also entail a complex hierarchical structure that is difficult to infer without supervision \cite{ssl6}.

Recently, many studies have demonstrated the empirical successes of SSL-based speech enhancement on low-resource clean speech data and highly reverberant environments. For example, T. Sun et al. propose a knowledge-assisted waveform framework (K-SENet) for speech enhancement \cite{icmla}. A perceptual loss function that relies on self-supervised speech representations pretrained on large datasets is used to provide guidance for the baseline network. Besides, H.-S. Choi et al. perturb information in the input signal and provide essential attributes for synthesis networks to reconstruct the input signal \cite{shouer}. Instead of using labels, a new set of analysis features is used, i.e., wav2vec feature and newly proposed pitch feature, Yingram, which allows for fully self-supervised training. However, both methods reply on large-scale training data, which is expensive to obtain. Therefore, the state-of-the-art SSL methods based on the limited training data are eager to develop.

\section{PROPOSED METHOD}
\subsection{Overall Architecture}

\begin{figure}[htbp!]
\centering
\includegraphics[width=8.5cm, height=8.5cm]{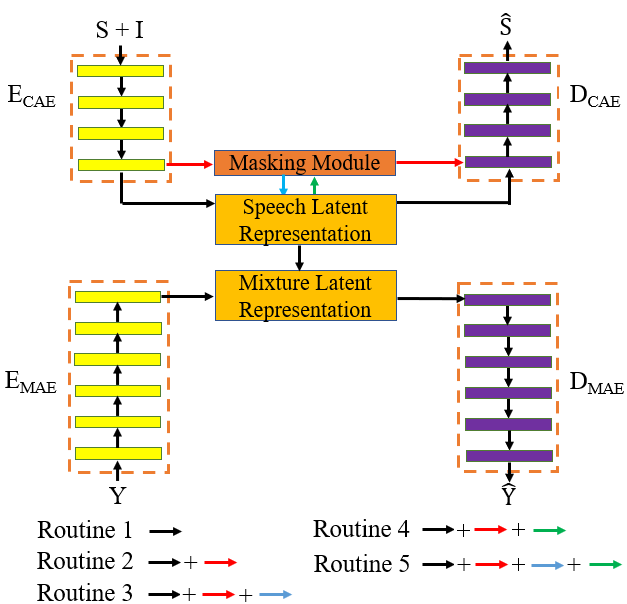}
\caption{The overall architecture of the proposed method. The clean speech $\mathbf{S}$ and interference $\mathbf{I}$ are fed into the $E_{CAE}$. The interference consists of background noises, reverberation of both speech and noise signals. After the feature combination is extracted, as the first pre-task, the latent representation of the clean speech signal is learned via $E_{CAE}$. As the second pre-task, the DM and ERM are estimated in the masking module. Besides, the proposed method utilizes the speech reconstruction losses of each pre-task to train the other pre-task. After the feature maps are recovered in the decoder, the reconstructed clean spectra are obtained as the output by using $D_{CAE}$. By using the learned speech representation into the mixture representation, the estimated mixtures are produced from the mixture autoencoder (MAE) with unpaired and unseen training mixture spectra $\mathbf{Y}$.}\centering
\end{figure}

The overall architecture is presented in Fig. 1. In the training stage, two variational autoencoders, one clean speech autoencoder (CAE) and one mixture autoencoder (MAE), are exploited for different tasks. The encoder and decoder of the CAE are denoted as $E_{CAE}$ and $D_{CAE}$, respectively. Similarly, $E_{MAE}$ and $D_{MAE}$ are used to present the encoder and decoder of the MAE respectively. Besides, in order to simplify the equations, $\mathbf{S}$, $\mathbf{I}$, and $\mathbf{Y}$ are used to replace $S(t, f)$, $I(t, f)$ and $Y(t, f)$, respectively.

The input of the $E_{CAE}$ consists of a limited set of clean speech signals, background noise, and reverberated both speech and noise signals. First, five features introduced in Related Work is extracted at the frame level and are concatenated with the corresponding delta features. Then, the encoder $E_{CAE}$ produces the latent representation of the clean speech signal by compressing the spectra into higher dimensional space. In the proposed method, two pre-tasks are considered for pre-training: latent representation learning and mask estimation. The first task aims to learn the latent representation of only clean speech signals by autoencoding on their magnitude spectra. In addtion, in the second task, DM and ERM are trained to describe the relationships from the target speech signal to the mixture as equations (4)$\&$(6). Both the latent representation and masks are trained by minimizing the discrepancy between the clean speech spectra and the corresponding reconstruction. The decoder is trained by the losses from two pre-tasks and use the estimated speech latent representation and estimated masks from pre-tasks to produce the target speech spectra as the output.
% Both $E_{CAE}$ and $D_{CAE}$ comprise 4 1-D convolutional layers. In the $E_{CAE}$, the size of the hidden dimension sequentially decreases from 512 $\rightarrow$ 256 $\rightarrow$ 128 $\rightarrow$ 64. Consequently, the dimension of the latent space is set to 64, and a stride of 1 sample with a kernel size of 7 for the convolutions. Different from $E_{CAE}$, $D_{CAE}$ increases the size of the latent dimensions inversely.

% It is highlighted that only is input to the first pre-task to facilitate the PAE to learn . Moreover, a limited set of paired examples of reverberant mixtures and the corresponding clean speech signals is available in the masking module.  The PAE network in the proposed method consists of 4 1-D convolutional layers in the encoder.

Different from the CAE, the MAE only needs to access the reverberant mixture. The $E_{MAE}$ obtains the reverberant mixture and extract the feature combination similar to $E_{CAE}$. Consequently, the latent representation of the mixture $\mathbf{X_{M}}$ is obtained as the output of $E_{MAE}$. The learned representation and masks from the CAE are exploited to modify the loss functions and learn a shared latent space between the clean speech and mixture representations. To achieve this, we use the CAE and incorporate the cycle-consistency terms into the overall loss. Then, two latent representations before and after the cycle loop through the CAE can be trained to be close. Benefited from the pre-tasks, a mapping function from the mixture spectra to the target speech spectra is learned with the latent representation of the clean speech signal. Furthermore, $D_{MAE}$ is trained to produce the estimated mixture as the downstream task.

% The MAE network follows a similar architecture to CAE. $E_{MAE}$ consists of 6 1-D convolutional layers where the hidden layer sizes decrease from 512 $\rightarrow$ 400 $\rightarrow$ 300 $\rightarrow$ 200 $\rightarrow$ 100 $\rightarrow$ 64, and $D_{MAE}$ increases the sizes inversely.

In the testing stage, because the loss function in $E_{MAE}$ is trained with the mapping of the latent space from the mixture spectra to the target speech spectra, the unseen reverberant mixtures are fed into the trained $E_{MAE}$ and the features are extracted. Then, the trained $E_{MAE}$ produces an estimated latent representation of the reverberant mixture. Finally, the trained $D_{CAE}$ obtains the reconstructed representation and maps to the target speech signal. 

\subsection{Feature Combination}
The feature plays an important part in the speech enhancement problem \cite{Student}. According to \cite{featureset}, different acoustic features characterize different properties of the speech signal. Therefore, we apply feature learning including spectrogram, MFCC, AMS, RASTP-PLP, and cochleagram which are commonly used in supervised speech enhancement to examine the performance of each feature in SSL. To achieve that, each of the five features is independently extracted from the spectra of clean speech signals and noisy mixtures. Then, each feature is severally used in the encoder to learn the latent representation. Besides, in the masking module, the DM and ERM are calculated with the feature combinations of the clean speech and the mixture spectra. Therefore, according to (7), the masks are applied to the reverberant mixture to estimate the target speech. Our feature learning study provides the different levels of speech enhancement performance improvement with different types of features.

Moreover, in order to further improve the speech enhancement performance compared to using the individual feature, feature combination is introduced to combine various complementary features \cite{Two}. A straightforward way of finding complementary features is to try all combinations of features. However, the number of combinations is exponential with respect to the number of features. Inspired by \cite{features}, group Lasso (least absolute shrinkage and selection operator) to quickly identify complementary features and the features that have relatively large responses are selected as the complementary features. After the features are extracted at the frame level and are concatenated with the corresponding delta features \cite{delta}. Then, the auto-regressive moving average (ARMA) filter is exploited to smooth temporal trajectories of all features \cite{filter}. Consequently, the feature combination based latent representation is used to estimate the loss between the clean speech and the reconstructed latent representations. The proposed SSL-based feature combination method is intuitive as it uses the complementary features in combination, and simple in that the selected features are estimated separately.
% Different from the conventional feature combination methods, the proposed weighting scheme provides a measure of the importance of each feature and is trained with the loss between the target and estimation, i.e., speech or mixture latent representation. For example, to exploit the proposed weighting scheme on the clean speech latent representation learning in CAE, first, the latent representation is extracted from each feature map. Then, the weight is assigned to each feature to generate the feature combination based latent representation as:
% % \begin{equation}
% % \mathbf{W_{S}}=\sum_{i=1}^{5} w_{S}(i)
% % \end{equation}

% \begin{equation}
% \mathbf{\hat{X}_{S}}= \sum_{i=1}^{5}w_{X_{S}}(i) \mathbf{\hat{X}_{S}}(i)
% \end{equation}
% where $\mathbf{\hat{X}_{S}}(i)$ is the estimated speech latent representation based on the $i$-th feature type. Besides, $i$ refers to the index of the feature types and $w_{X_{S}}(i)$ denotes the $i$-th weight for the clean speech latent representation. In the proposed method, $i$ is ranged between 1-5 because five features are exploited to construct the feature maps. 

\subsection{Ensemble Pre-Tasks}
Different from the single pre-task SSL methods, the proposed method exploits the masking module to further improve the denoising and dereverberation performances. In this work, the internal effectiveness between the two pre-tasks is studied. Therefore, we design five routines to differently train the models with the same input.

Routine 1 uses the single pre-task as \cite{ssl}. The proposed masking module is introduced as the second pre-task in the routine 2. Moreover, the routine 3 applies the loss from latent representation learning to help train the masking module, while vice versa in the routine 4. Finally, the The losses from each pre-task is used to train the other one in the routine 5.
\subsubsection{Routine 1}
The original single pre-task method similar to \cite{ssl} is used in this routine. A limited training set of clean speech signals are exploited to learn the latent representation and a mapping from the mixture to the target speech is learned with the latent representation of the desired speech signal.

We use two loss terms to calculate the overall loss for the CAE. The discrepancy between the clean speech spectra and the reconstruction $\hat{\mathbf{S}}$ with the $L$2 norm of the error is calculated as:
\begin{equation}
\mathcal{L}_{\mathbf{S}}=\|\mathbf{S}-\hat{\mathbf{S}}\|_{2}^{2}
\end{equation}
The Kullback-Leibler (KL) loss of the CAE is denoted as $\mathcal{L}_{\text {KL-CAE }}$ and is applied to train the latent representation closed to a normal distribution \cite{ssl}. Therefore, the overall for the CAE is given as:
\begin{equation}
\mathcal{L}_{\text {CAE}}=\lambda_{1}  \cdot\mathcal{L}_{\text {KL-CAE }}+\mathcal{L}_{\mathbf{S}}
\end{equation} 
The coefficient $\lambda_{1}$ is added and set to 0.001. Similarly, the $\mathcal{L}_{\mathbf{Y}}$ denotes the loss between the noisy mixture and the corresponding reconstruction $\hat{\mathbf{Y}}$ as:
\begin{equation}
\mathcal{L}_{\mathbf{Y}}=\|\mathbf{Y}-\hat{\mathbf{Y}}\|_{2}^{2}
\end{equation}
Besides, in order to enforce a shared latent representation between the two autoencoders, the mixture cycle loss $\mathcal{L}_{\text {Y-cyc }}$ is added as:
\begin{equation}
\mathcal{L}_{\text {Y-cyc }}=\|\mathbf{Y}-\hat{\mathbf{Y}}\|_{2}^{2}+\lambda_{2} \cdot\left\|\mathbf{X_{Y}}-\hat{\mathbf{X}}_{\mathbf{Y}}\right\|_{2}^{2}
\end{equation}
where $\lambda_{2}=0.01$. $\mathbf{X_{Y}}$ and $\hat{\mathbf{X}}_{\mathbf{Y}}$ denote the latent representation of noisy mixture and the reconstruction, respectively. The latent representation is fed into the MAE decoder for mapping the target speech spectrogram from the mixture spectrogram. Then, the mapping representation helps the CAE to obtain the reconstruction. The input mixture spectrogram is resembled by the cycle reconstruction of the mixture spectrogram. Besides, the two latent representations are close with the CAE losses. Furthermore, the overall loss to train the MAE is a combination of loss terms with the KL loss $\mathcal{L}_{\text {KL-MAE }}$ as:
\begin{equation}
% \mathcal{L}_{\text {masking }}=\mathcal{L}_{\mathbf{S}}+\mathcal{L}_{\text {cyc }}+\mathcal{L}_{\mathbf{I}}+\lambda \cdot \mathcal{L}_{\text {KL-MAE }}
\mathcal{L}_{\text {MAE}}=\lambda_{3}  \cdot \mathcal{L}_{\text {KL-MAE }}+\mathcal{L}_{\mathbf{Y}}+\mathcal{L}_{\text {Y-cyc }}
\end{equation}
where $\lambda_{3}$ is the coefficient of $\mathcal{L}_{\text {KL-masking }}$ and empirically set to 0.001. In the testing stage, the path $E_{MAE} \rightarrow D_{CAE}$ provides the estimated speech. However, speech enhancement performance of the routine 1 is limited due to the single pre-task. Therefore, the second pre-task is introduced to improve performance in the routine 2.
\subsubsection{Routine 2} 
Compared to the routine 1, the second pre-task is added in the routine 2. The pre-tasks are designed in parallel between the $E_{CAE}$ and $D_{CAE}$. After the feature combinations are extracted, the latent representation is obtained from the first pre-task and the masking module obtains the feature combination to produce the estimated speech as the second pre-task. The architecture of the masking module is depicted in Fig. 2.
\begin{figure}[htbp!]
\centering
\includegraphics[width=4cm, height=8.5cm]{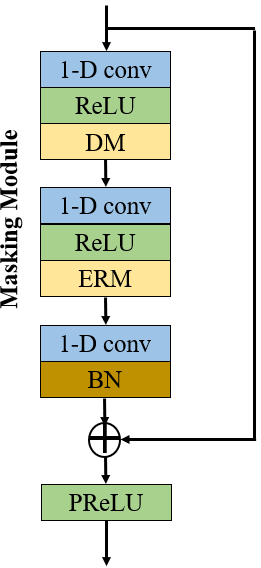}
\caption{The architecture of the proposed masking module. The feature combinations of the clean speech, interference, and noisy mixture are fed into the masking module. After the DM and ERM are estimated, the estimated speech are produced for the decoder.}\centering
\end{figure}

The masking module has three sub-layers and aims to estimate the clean speech feature combination. To achieve this, the first two sub-layers consists of two T-F masks, DM and ERM, respectively. After the feature combinations of speech signals, interferences and noisy mixtures are obtained from the first sub-layer, 1D convolutional layers with a kernel size of 1 $\times$ 7 are used to enlarge the receptive field along the frequency axis \cite{receptive}. Then, the DM is applied to model the relationship between the dereverberated mixture and the noisy mixture as (4). However, the dereverberation with only the DM is very challenging in a highly reverberant scenario \cite{IET}. Therefore, in the second sub-layer, the ERM is used to better estimate the relationship between the clean speech and the estimated dereverberated mixture as (6). Both sub-layers are followed by batch normalization (BN) to accelerate the model training \cite{bn}. The estimated speech feature combination $\hat{\mathbf{S}}^{\prime}$ from the masking module can be obtained with the sequentially trained sub-layers as the multiplication of estimated masks. The losses from two pre-tasks, i.e., latent representation learning and masking module, are jointly train the $D_{CAE}$ to estimate the final clean speech.

In the downstream training, the unseen and unpaired noisy mixture spectra are fed into the MAE and the feature combination is extracted from the spectra. Different from the upstream training, we only consider one way to reconstruct the mixture spectra. First, the noisy mixture is encoded in the $E_{MAE}$. At the bottleneck of the MAE, on the one hand, the latent representation of the noisy mixture is learned. On the other hand, the mixture cycle loss $\mathcal{L}_{\text {Y-cyc }}$ is added to enforce the shared latent space between two autoencoders as (11). Consequently, the estimated mixture latent representation can be generated. At the final step, the $D_{MAE}$ produce the final estimated mixture. According to the routine 2, the target speech is estimated with the ensemble pre-tasks. However, the estimations from each pre-task have different levels of degradation compared to the clean speech. Therefore, in the routines 3$\&$4, the loss from one pre-task is used to train the other.

\subsubsection{Routine 3} As aforementioned, in the routine 3, the learned latent representation is further used to train the masking module. We first calculate the temporal masking module loss as:
\begin{equation}
\mathcal{L}_{\mathbf{S}^{\prime}_{masking}}=\left\|\mathbf{S}-\hat{\mathbf{S}}^{\prime}\right\|_{2}^{2}
\end{equation}
where $\prime$ denotes temporal terms. In the first pre-task, the latent representation of the clean speech is learned by minimizing the loss between the clean latent representation $\mathbf{X_{S}}$ and the reconstruction $\hat{\mathbf{X}}_{\mathbf{S}}$ as:
\begin{equation}
\mathcal{L}_{\mathbf{X_{S}}}=\left\|\mathbf{X_{S}}-\hat{\mathbf{X}}_{\mathbf{S}}\right\|_{2}^{2}
\end{equation}
Then, the latent space loss $\mathcal{L}_{\mathbf{X_{S}}}$ is added to further minimize the masking module loss as:
\begin{equation}
\mathcal{L}_{\mathbf{S}_{masking}^{r3}}=\left\|\mathbf{S}-\hat{\mathbf{S}}^{\prime}\right\|_{2}^{2}+\lambda_{4}  \cdot\mathcal{L}_{\mathbf{X_{S}}}
\end{equation}
where $r3$ denotes the routine 3. The coefficient $\lambda_{4}$ is added as a constraint and set to 0.1. After the masking module loss is minimized, the overall loss to train the CAE can be calculated as:
\begin{equation}
% \mathcal{L}_{\text {masking }}=\mathcal{L}_{\mathbf{S}}+\mathcal{L}_{\text {cyc }}+\mathcal{L}_{\mathbf{I}}+\lambda \cdot \mathcal{L}_{\text {KL-MAE }}
\mathcal{L}_{\text {CAE}}^{r3}=\lambda_{5}  \cdot \mathcal{L}_{\text {KL-CAE }}+\mathcal{L}_{\mathbf{S}}+\mathcal{L}_{\mathbf{S}_{masking}}+\mathcal{L}_{\mathbf{X_{S}}}
\end{equation}
where $\lambda_{5}$ is the coefficient of $\mathcal{L}_{\text {KL-masking }}$ and empirically set to 0.001. After the MAE is trained, the estimated speech can be obtained from the path $E_{MAE}\rightarrow D_{CAE}$.

\subsubsection{Routine 4} Different from the routine 3, the output from the masking module helps to learn the target latent presentation in the routine 4. Firstly, the temporal latent representation loss is calculated as:
\begin{equation}
\mathcal{L}_{\mathbf{X_{S}^{\prime}}}=\left\|\mathbf{X_{S}}-\hat{\mathbf{X}}_{\mathbf{S}}^{\prime}\right\|_{2}^{2}
\end{equation}
In the second pre-task, the masking module is trained to estimate the clean speech by minimizing the loss between the clean speech and the temporal reconstruction as:
\begin{equation}
\mathcal{L}_{\mathbf{S}_{masking}^{r4}}=\left\|\mathbf{S}-\hat{\mathbf{S}}^{\prime}\right\|_{2}^{2}
\end{equation}
Then, the masking module loss $\mathcal{L}_{\mathbf{S}_{masking}^{r4}}$ is added to improve the estimation accuracy of the clean speech latent representation with the loss term as:
\begin{equation}
\mathcal{L}_{\mathbf{X_{S}}}=\left\|\mathbf{X_{S}}-\hat{\mathbf{X}}_{\mathbf{S}}^{\prime}\right\|_{2}^{2}+\lambda_{6}  \cdot\mathcal{L}_{\mathbf{S}_{masking}^{r4}}
\end{equation}
where the coefficient $\lambda_{6}$ is set to 0.1. The overall loss of the CAE is similar to the routine 3 as (16). Compared to the routine 2, the latent representation is better estimated with the estimation from the masking module. In the downstream task training, the further trained latent representation helps the MAE to improve the noisy mixture estimation with the mixture cycle loss.

\subsubsection{Routine 5} In order to further improve the speech enhancement performance, the routines 3$\&$4 are combined in the routine 5. The losses from each pre-task are exploited to further train the other in the CAE. In the testing stage, the path $E_{MAE}\rightarrow D_{CAE}$ provides the estimated speech. The pseudocode of the CAE training in the routine 5 is summarized as Algorithm 1.
% The configuration difference of various routines is shown in TABLE I.
% \begin{table}[htbp!]
% \centering
% \caption{comparison of ssl methods with the proposed approach. the PT-FT method use 50800 paired utterances in the training stage, however, only 200 in the proposed method. moreover, the number of pre-tasks is set to 1, 1, 3, and 2 in three methods, respectively.}
% %\small\addtolength{\tabcolsep}{-1pt}
% \begin{tabular}{c|c|c|c|c|c}
% \hline
% Routine &1 & 2  & 3 &4&5 \\    
%  \hline
% % Unprocessed  & 41.5&1.44 &3.04 & \\
%  Single Pre-Task& \checkmark& \xmark & \checkmark  & \checkmark    \\
% \hline
% Paired Data & \checkmark & \xmark & \checkmark & \checkmark    \\
% \hline
%   Multiple Models &\xmark  &\checkmark&\xmark & \checkmark\\
%  \hline 
%   Single Pre-Task &\checkmark &\checkmark &\xmark & \xmark\\
%  \hline 
%   Reverberation &\xmark &\checkmark &\xmark &\checkmark\\
%  \hline 
% \end{tabular}
% \end{table}

\makeatletter
\newcommand{\AlgoResetCount}{\renewcommand{\@ResetCounterIfNeeded}{\setcounter{AlgoLine}{0}}}
\newcommand{\AlgoNoResetCount}{\renewcommand{\@ResetCounterIfNeeded}{}}
%\newcounter{AlgoSavedLineCount}
%\newcommand{\AlgoSaveLineCount}{\setcounter{AlgoSavedLineCount}{\value{AlgoLine}}}
%\newcommand{\AlgoRestoreLineCount}{\setcounter{AlgoLine}{\value{AlgoSavedLineCount}}}
%\RestyleAlgo{boxed}% Change from the 'ruled' style to 'boxed'
\SetAlgoNoLine% Removes 'vlined' option (somewhat opposite of \SetAlgoVlined)
\LinesNumberedHidden% Removes 'linesnumbered' option (opposite of \LinesNumbered)\qquad //mini-batch mean 
%\AlgoSaveLineCount% Stores the algorithm line number (similar to 'resetcount' in the package load option)
\makeatother

\begin{algorithm}

  \SetKwInOut{Input}{input}\SetKwInOut{Output}{output}

  \Input{Clean spectra $\mathbf{S}$, interferences $\mathbf{I}$, noisy speech spectra $\mathbf{Y}$, learning rate $\eta$, epoch $E_{max}$}
  \Output{Estimated clean speech $\hat{\mathbf{S}}$}
%   \BlankLine
  Initialize CAE and MAE parameters\;
  \For{$E = 1, 2, ..., E_{max}$  }{
   
    % $\mathbf{Z}_{1}$, $Real(\mathbf{S_{1}})$, $Imag(\mathbf{S_{1}})$ $\leftarrow$ $E_{1}(\mathbf{Y}_{1})$ \;
    
     Obtain the latent representations $\mathbf{X_{S}}\leftarrow$ $\mathbf{S}$ and $\mathbf{X_{Y}}$ $\leftarrow$ $\mathbf{Y}$\; 
    %   // Train the $E_{CAE}$ with the feature combination\;  
      
    Calculate $\mathcal{L}_{\mathbf{X_{S}^{\prime}}}$ $\leftarrow$ $\mathbf{X_{S}}$, $\hat{\mathbf{X}}_{\mathbf{S}}^{\prime} \qquad$ // First Pre-Task\;
    Estimate the $\widehat{DM}$ and $\widehat{ERM}$ $\leftarrow$ $\mathbf{S}$, $\mathbf{I}$, and $\mathbf{Y}$\;
    Estimate $\hat{\mathbf{S}}^{\prime}$ $\leftarrow$ $\mathbf{Y}$, $\widehat{DM}$, and $\widehat{ERM}$\;
    Calculate $\mathcal{L}_{masking}^{\prime} \qquad$ // Second Pre-Task\;
    Update the $\hat{\mathbf{X}}_{\mathbf{S}}$ $\leftarrow$ $\mathcal{L}_{\mathbf{X_{S}^{\prime}}}$, $\mathcal{L}_{masking}^{\prime}$\;
    Update the $\hat{\mathbf{S}}$ $\leftarrow$ $\mathcal{L}_{\mathbf{X_{S}^{\prime}}}$, $\mathcal{L}_{masking}^{\prime}$\;
    Train CAE by minimizing $\mathcal{L}_{\text {CAE}}$\;
 }
  \caption{Routine 5 pesudocode.}\label{algo_disjdecomp}
\end{algorithm}

% The configuration difference of various routines is shown in TABLE I.
% \begin{table}[htbp!]
% \centering
% \caption{comparison of ssl methods with the proposed approach. the PT-FT method use 50800 paired utterances in the training stage, however, only 200 in the proposed method. moreover, the number of pre-tasks is set to 1, 1, 3, and 2 in three methods, respectively.}
% %\small\addtolength{\tabcolsep}{-1pt}
% \begin{tabular}{c|c|c|c|c|c}
% \hline
% Routine &1 & 2  & 3 &4&5 \\    
%  \hline
% % Unprocessed  & 41.5&1.44 &3.04 & \\
%  Single Pre-Task& \checkmark& \xmark & \checkmark  & \checkmark    \\
% \hline
% Paired Data & \checkmark & \xmark & \checkmark & \checkmark    \\
% \hline
%   Multiple Models &\xmark  &\checkmark&\xmark & \checkmark\\
%  \hline 
%   Single Pre-Task &\checkmark &\checkmark &\xmark & \xmark\\
%  \hline 
%   Reverberation &\xmark &\checkmark &\xmark &\checkmark\\
%  \hline 
% \end{tabular}
% \end{table}

\section{EXPERIMENTAL RESULTS}
\subsection{Comparisons}
The proposed method is compared with three state-of-the-art SSL speech enhancement approaches \cite{ssl, cl, ssl1} on two publicly-available datasets. The first method is SSE \cite{ssl} which exploits two autoencoders to process pre-task and downstream task, respectively. The second method is pre-training fine-tune (PT-FT) \cite{ssl1}, which uses three models and three SSL approaches for pre-training: speech enhancement, masked acoustic model with alteration (MAMA) used in TERA \cite{mama} and continuous contrastive task (CC) used in wav2vec 2.0 \cite{wav}. The PT-FT method is reproduced with DPTNet model \cite{DT} and three pre-tasks because it shows the best speech enhancement performance in \cite{ssl1}. The third method applies a simple contrastive learning (CL) procedure which treats the abundant noisy data as makeshift training targets through pairwise noise injection \cite{ssl1}. In the baseline, the recurrent neural network (RNN) outputs with a fully-connected dense layer with sigmoid activation to estimate a time-frequency mask which is applied onto the noisy speech spectra. The configuration difference is shown in TABLE I. The cross mark \xmark$\ $means the method does not use the setting such as no reverberations in \cite{cl} but does not mean it cannot be handled in the method.
\FloatBarrier
\begin{table}[h]
\centering
\caption{Comparison of SSL methods with the proposed approach. The PT-FT method use 50,800 paired utterances in the training stage, however, only 200 unpaired utterances are applied in the proposed method. Moreover, the number of pre-tasks is set to 3 and 2 in the PT-FT and proposed method, respectively.}
\small\addtolength{\tabcolsep}{-1pt}
\begin{tabular}{c|c|c|c|c}
\hline
 & CL \cite{cl}& SSE \cite{ssl}  & PT-FT \cite{ssl1} &\textit{Proposed } \\    
 \hline
% Unprocessed  & 41.5&1.44 &3.04 & \\
Noise & \checkmark& \xmark & \checkmark  & \checkmark    \\
\hline
Paired Data & \checkmark& \xmark & \checkmark & \xmark    \\
\hline
  Multiple Models &\xmark  &\checkmark&\xmark & \checkmark\\
 \hline 
  Single Pre-Task &\checkmark &\checkmark &\xmark &\xmark\\
 \hline 
  Reverberation  &\xmark &\checkmark &\xmark &\checkmark\\
 \hline 
\end{tabular}
\end{table}
\FloatBarrier

\begin{table*}[htbp!]
\caption{speech enhancement performance in terms of three noise interferences at four snr levels in ipad$\char`_$livingroom1. each result is the average value of 600 experiments. \textit{italic} shows the proposed methods. {\bfseries bold} indicates the best result.}
\centering
%\small\addtolength{\tabcolsep}{-1pt}
\begin{tabular}{ccccccccccccccccc}
\hline
& \multicolumn{4}{c}{PESQ} & %
    \multicolumn{4}{c}{CSIG} & \multicolumn{4}{c}{CBAK}& \multicolumn{4}{c}{COVL}\\
\cline{2-17}
SNR (dB)  &-10&-5 & 0 & 5 &-10 &-5 & 0 & 5 &-10 &-5 & 0 & 5 &-10 &-5 & 0 & 5 \\
CL \cite{cl}   &1.43 &1.52&1.54&1.60 &1.96&2.20&2.30&2.40 & 1.57& 1.76&1.92&2.03&1.55&1.77&1.86&1.94 \\
 SSE \cite{ssl}   &1.48&1.53&1.56&1.58 &2.04&2.30&2.39&2.45 & 1.63& 1.83&1.94&2.10&1.68&1.81&1.88&2.00 \\
PT-FT \cite{ssl1} &1.52&1.55&1.59&1.62 &2.10&2.28&2.34&2.43 & 1.67& 1.81&1.96&2.08&1.68&1.78&1.89&2.00 \\

\textit{Proposed}   &{\bfseries 1.74}&{\bfseries 1.80}&{\bfseries 1.83}&{\bfseries 1.89}&{\bfseries 2.47} &{\bfseries 2.56}&{\bfseries 2.59}&{\bfseries 2.63} & {\bfseries 1.95}& {\bfseries 2.02}& {\bfseries 2.15}& {\bfseries 2.30}& {\bfseries 1.86}& {\bfseries 1.98}& {\bfseries 2.03}& {\bfseries 2.17}\\
 \hline 
\end{tabular}
\end{table*}

\begin{table*}[htbp!]
\caption{speech enhancement performance in terms of three noise interferences at four snr levels in ipad$\char`_$bedroom1. each result is the average value of 600 experiments. \textit{italic} shows the proposed methods. {\bfseries bold} indicates the best result.}
\centering
%\small\addtolength{\tabcolsep}{-1pt}
\begin{tabular}{ccccccccccccccccc}
\hline
& \multicolumn{4}{c}{PESQ} & %
    \multicolumn{4}{c}{CSIG} & \multicolumn{4}{c}{CBAK}& \multicolumn{4}{c}{COVL}\\
\cline{2-17}
SNR (dB)  &-10&-5 & 0 & 5 &-10 &-5 & 0 & 5 &-10 &-5 & 0 & 5 &-10 &-5 & 0 & 5 \\
CL \cite{cl}   &1.45 &1.57&1.59&1.61&1.93 &2.25&2.32&2.39& 1.69 & 1.82&1.99&2.08&1.70&1.82&1.90&2.03  \\
 SSE \cite{ssl}  &1.50&1.59&1.62&1.65&2.11 &2.34&2.43&2.49& 1.72 & 1.88&1.97&2.16&1.73&1.84&1.89&2.02 \\
PT-FT \cite{ssl1}  &1.57&1.64&1.73&1.74 &2.16&2.33&2.46&2.51 & 1.75& 1.91&2.03&2.19&1.77&1.85&1.94&2.05 \\

\textit{Proposed}   &{\bfseries 1.82}&{\bfseries 1.91}&{\bfseries 1.96}&{\bfseries 2.05}&{\bfseries 2.47} &{\bfseries 2.58}&{\bfseries 2.64}&{\bfseries 2.69} & {\bfseries 1.99}& {\bfseries 2.08}& {\bfseries 2.20}& {\bfseries 2.31}& {\bfseries 1.93}& {\bfseries 2.02}& {\bfseries 2.11}& {\bfseries 2.17}\\
 \hline 
\end{tabular}
\end{table*}

\begin{table*}[htbp!]
\caption{speech enhancement performance in terms of three noise interferences at four snr levels in ipad$\char`_$confroom1. each result is the average value of 600 experiments. \textit{italic} shows the proposed methods. {\bfseries bold} indicates the best result.}
\centering
%\small\addtolength{\tabcolsep}{-1pt}
\begin{tabular}{ccccccccccccccccc}
\hline
& \multicolumn{4}{c}{PESQ} & %
    \multicolumn{4}{c}{CSIG} & \multicolumn{4}{c}{CBAK}& \multicolumn{4}{c}{COVL}\\
\cline{2-17}
SNR (dB)  &-10&-5 & 0 & 5 &-10 &-5 & 0 & 5 &-10 &-5 & 0 & 5 &-10 &-5 & 0 & 5 \\
 CL \cite{cl} &1.48&1.58&1.62&1.63 &2.09&2.26&2.33&2.44 & 1.77& 1.84&2.00&2.09&1.81&1.85&1.92&2.06  \\
 SSE \cite{ssl}  &1.53&1.61&1.65&1.66 &2.12&2.35&2.46&2.47 & 1.78& 1.93&2.00&2.17&1.80&1.85&1.90&2.05 \\
PT-FT \cite{ssl1}   &1.60&1.66&1.74&1.77 &2.18&2.34&2.45&2.53 & 1.83& 1.94&2.05&2.23&1.96&\textbf{2.02}& 2.07&2.10 \\

\textit{Proposed}   &{\bfseries 1.85}&{\bfseries 1.97}&{\bfseries 2.00}&{\bfseries 2.04}&{\bfseries 2.48} &{\bfseries 2.60}&{\bfseries 2.65}&{\bfseries 2.72} & {\bfseries 2.05}& {\bfseries 2.13}& {\bfseries 2.21}& {\bfseries 2.36}& {\bfseries 2.01} &\textbf{2.02}& {\bfseries 2.15}& {\bfseries 2.25}\\
 \hline 
\end{tabular}
\end{table*}
\subsection{Datasets}
In the CAE training, 12 clean utterances from 4 different speakers with three reverberant room environments (ipad$\char`_$livingroom1, ipad$\char`_$bedroom1, and ipad$\char`_$confroom1) are randomly selected from the DAPS dataset \cite{daps}. The training data consists of 2 male and 2 female speakers each reading out 3 utterances and recorded in different indoor environments with different real room impulse responses (RIRs) \cite{daps}. In the MAE training, the unseen and independent 300 noisy mixtures from 20 different speakers with three reverberant room environments are randomly selected from the DAPS dataset. The training data consists of 10 male and 10 female speakers each reading out 5 utterances and recorded in different indoor environments with different real room impulse responses (RIRs) \cite{daps}. In order to improve the ability of the proposed method in adapting to unseen speakers, the speakers in the MAE training are manually designed to be different from the speakers in the CAE training. Moreover, three background noises ($factory$, $babble$, and $cafe$) from the NOISEX dataset \cite{noise} and four SNR levels (-10, -5, 0, and 5 dB) are used to generate the mixtures in both the CAE and MAE. The validation data contains 50 noisy mixtures generated by the randomly selected reverberant speech from the DAPS dataset and the background noise. In the testing stage, 200 reverberant utterances of 10 speakers are randomly selected and used to generate the mixtures with the same background noises and SNR levels for the configuration in the training stage. Therefore, the numbers of mixtures in CAE training, MAE training, validation and testing data are 432, 10, 800, 1, 800 and 7, 200, respectively. It is highlighted that the speakers in the training and testing stages are unseen in proposed method and all baselines.

%   .  In each environment, 12 utterances are randomly selected to generate the training data to train the CAE. Then, the rest unseen and independent 188 utterances are exploited for the MAE to obtain the estimated mixtures. Therefore, the training data in the CAE and MAE is unseen and unpaired. The validation data contains 50 noisy mixtures generated by the randomly selected reverberant speech from the DAPS dataset and the background noise. In the testing stage, 200 reverberant utterances of 10 speakers are randomly selected and used to generate the mixtures with the same background noises and SNR levels for the configuration in the training stage. 

\subsection{Experiment Setup}
Both $E_{CAE}$ and $D_{CAE}$ comprise 4 1-D convolutional layers. In the $E_{CAE}$, the size of the hidden dimension sequentially decreases from 512 $\rightarrow$ 256 $\rightarrow$ 128 $\rightarrow$ 64. Consequently, the dimension of the latent space is set to 64, and a stride of 1 sample with a kernel size of 7 for the convolutions. Different from $E_{CAE}$, $D_{CAE}$ increases the size of the latent dimensions inversely.

The MAE network follows a similar architecture to CAE. $E_{MAE}$ consists of 6 1-D convolutional layers where the hidden layer sizes decrease from 512 $\rightarrow$ 400 $\rightarrow$ 300 $\rightarrow$ 200 $\rightarrow$ 100 $\rightarrow$ 64, and $D_{MAE}$ increases the sizes inversely.

The proposed method is trained by using the Adam optimizer with a learning rate of 0.001 and the batch size is 20. The number of training epochs for CAE and MAE are 700 and 1500, respectively. All the experiments are run on a workstation with four Nvidia GTX 1080 GPUs and 16 GB of RAM. The complex speech spectra have 513 frequency bins for each frame as a Hanning window and a discrete Fourier transform (DFT) size of 1024 samples are applied.

According to \cite{ssl}, we use composite metrics that approximate the Mean Opinion Score (MOS) including COVL: MOS predictor of overall processed speech quality, CBAK: MOS predictor of intrusiveness of background noise, CSIG: MOS predictor of signal distortion \cite{mos} and Perceptual Evaluation of Speech Quality (PESQ). Besides, the signal-to-distortion ratio (SDR) is evaluated in terms of baselines and the proposed method. Higher values of the measurements imply better enhancement performance.

\subsection{Comparison with SSL methods}
The speech enhancement performance of the proposed method with the routine 5 and feature combination is compared with state-of-the-art SSL methods in TABLES I-III.

It can be seen from TABLES II-IV that the proposed method outperforms the state-of-the-art SSL methods in terms of all three performance measures. The proposed method has 16.1$\%$, 16.5$\%$, and 18.7$\%$ improvements compared with the PT-FT method in terms of PESQ at -5 dB SNR level in three environments. The environment ipad$\char`_$livingroom1 is relatively more reverberant compared to the other two rooms \cite{daps}, while the improvement in performance is still significant. For example, in TABLE I, the proposed method has 13.3$\%$, 9.5$\%$, and 10.6$\%$ improvements compared to the CL, SSE, and PT-FT methods in terms of CBAK at 5 dB, respectively. Besides, speech enhancement comparisons at four different SNR levels are shown in TABLES I-III. From the experimental results, the performance improvement compared to the baselines is obvious even at relatively low SNR level i.e., -10 dB. Compared to the PT-FT method, the proposed method has 10.7$\%$, 11.2$\%$, 7.4$\%$ and 8.5$\%$ improvements in terms of COVL at four SNR levels.

\begin{figure*}[htbp!]
\centering
\includegraphics[width=17.5cm, height=5cm]{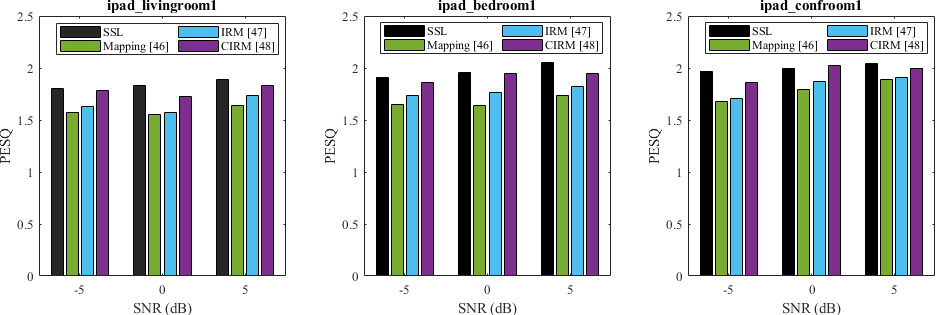}
\caption{Comparisons with supervised learning-based methods at three SNR levels in three environments (ipad$\char`_$livingroom1, ipad$\char`_$bedroom1, and ipad$\char`_$confroom1). Each result is the average value of 200 experiments.}\centering
\end{figure*}
In \cite{ssl1}, the original PT-FT method is trained with Libri1Mix train-360 set \cite{lib} which contains 50,800 utterances. However, in the comparison experiments, we use the limited amount of training utterances (200). Therefore, the speech enhancement performance of the PT-FT suffers a significant degradation compared with the original implementation. Moreover, the speech enhancement performance of each feature is limited. However, the proposed method takes advantage of each feature in the feature combination and addresses the speech enhancement problem. Thus, the speech enhancement performance is improved compared with only extracting one type of feature from the clean speech representation in the SSL methods.
\begin{table*}[htbp!]
\caption{feature ablation studies in terms of three noise interferences ($factory$, $babble$, and $cafe$) at three SNR levels in ipad$\char`_$confroom1. each result is the average value of 600 experiments. the proposed masking module and ensemble learning are available in all five implementations. }
\centering
%\small\addtolength{\tabcolsep}{-1pt}
\begin{tabular}{ccccccccccccc}
\hline
\multirow{2}{*}{Feature} & \multicolumn{3}{c}{PESQ} & %
    \multicolumn{3}{c}{CSIG} & \multicolumn{3}{c}{CBAK}& \multicolumn{3}{c}{COVL} \\
\cline{2-13}
 &-5 dB& 0 dB& 5 dB&-5 dB& 0 dB& 5 dB&-5 dB& 0 dB& 5 dB&-5 dB& 0 dB& 5 dB\\
Spectrogram \cite{ssl}  &1.61&1.65&1.66 &2.35&2.46&2.47 & 1.93&2.00&2.17&1.85&1.90&2.05 \\
MFCC \cite{mfcc}   &1.67&1.70&1.73 &2.44&2.47&2.52 & 2.01&2.11&2.24&1.97&2.02&2.10 \\
AMS \cite{AMS2}   &1.81&1.84&1.89 &2.52&2.61&2.67 & 2.09&2.18&2.30&2.00&2.11&2.19 \\
RASTA-PLP \cite{PLP2}   &1.62&1.65&1.72 &2.38&2.50&2.51 & 1.97&2.07&2.19&1.92&2.03&2.05 \\
cochleagram   &1.66&1.67&1.71 &2.39&2.48&2.50 & 1.97&2.06&2.21&1.95&1.94&2.07 \\
 \hline 
\end{tabular}
\end{table*}

In the proposed method, the mismatch of the speakers between the training and testing stages is solved, which is most important in practical scenarios e.g., speaker-independent. Moreover, the proposed method can be used where both SNR levels and noise types are unseen, however, the speech enhancement performance suffers a slight degradation, which will be handled in future work.
\subsection{Comparison with SL methods}
Recently, most of speech enhancement methods are developed based on supervised learning (SL) due to the promising performance under the sufficient training data. However, in the practical scenarios, the training frequently suffers the problem which lacking in paired data. Therefore, in order to show the competitiveness of the proposed SSL method, the mapping- and masking-based supervised methods are reproduced with the same number of training data \cite{mapping, IRM1, cirm}. The SL baselines are implemented with deep neural networks (DNNs) which use three hidden layers, each having 1024 rectified linear hidden units as the original implementations. Apart form the ideal ratio mask (IRM), we also compare the proposed phase-aware method with the complex ideal ratio mask (cIRM). The experimental results of comparisons with the SL methods are presented in Fig. 3. The SL methods are evaluated with the unseen speakers as the proposed method.
%jixu
In Fig. 3, we can observe that the proposed SSL method shows better performance than the SL methods. On the one hand, different from the original experimental settings \cite{mapping, IRM1, cirm}, the SL methods are evaluated in a challenging scenario with highly reverberant environments, limited training data, and unseen speakers, which suffers a significant performance degradation. However, the proposed SSL method solves the limitations such as the mismatch between the training and testing conditions to guarantee the speech enhancement performance. On the other hand, the compared baselines are not state-of-the-art approaches. However, the SSL research in speech enhancement problem just started \cite{ssl}. We simply provide the comparison between the SSL and SL study to show the competitiveness of the proposed SSL method. Besides, the experiments are set up in a challenging indoor environments with high reverberations as the practical scenarios. Therefore, the speech enhancement of the proposed and baseline SSL methods is relatively less than the SL methods with the advantage of SSL methods to be used in practical scenarios e.g., the mixtures are only available in real room recordings.
\subsection{Ablation Study}
Firstly, the effectiveness of each feature is investigated in the SSL study. It is highlighted that the proposed masking module and ensemble learning are not introduced in this comparison. The experimental results with four SNR levels (-10, -5, 0, and 5dB) and three room environments (ipad$\char`_$livingroom1, ipad$\char`_$bedroom1, and ipad$\char`_$confroom1) are shown in TABLE V.

From TABLE V, it can be observed that AMS outperforms the other four features in terms of four performance measures. For example, the AMS based SSL method has 8.4$\%$, 7.7$\%$, 11.7$\%$ and 9.1$\%$ improvements compared to the other four features in terms of PESQ at -5 dB SNR level. As proven in the previous study, AMS mimics important aspects of the human auditory system in combination with mel-scale cepstral coefficients \cite{ams4}. The experimental results show the speech enhancement performance of various features in SSL study and provide each contribution of each feature in the proposed feature combination method.

In order to study the effectiveness of ensemble learning, the averaged speech enhancement performance of five routines with four SNR levels (-10, -5, 0, and 5dB) and three room environments (ipad$\char`_$livingroom1, ipad$\char`_$bedroom1, and ipad$\char`_$confroom1) are compared in TABLE VI with the feature combination.

\FloatBarrier
\begin{table}[h]
\caption{ablation study of five training routines of three snr levels (-5, 0, and 5 dB), three noise interferences ($factory$, $babble$, and $cafe$) in ipad$\char`_$livingroom1. each result is the average value of 1800 experiments.}
\centering
%\small\addtolength{\tabcolsep}{-1pt}
\begin{tabular}{cccccccc}
\hline
Routine& PESQ & CSIG  & CBAK &COVL &SDR (dB) \\    
 \hline
% Unprocessed  & 41.5&1.44 &3.04 & \\
1  & 1.56& 2.39 & 1.94  & 1.88  & 5.16 \\
\hline
2 & 1.80 & 2.56 & 2.17 & 2.02 &  6.94 \\
\hline
  3 &1.86  &2.61&2.18 & 2.06 &8.19 \\
 \hline 
  4 &1.84 &2.59 &2.18 & 2.04&7.88 \\
 \hline 
  5 &{\bfseries 1.94} &{\bfseries 2.63} &{\bfseries 2.20} &{\bfseries 2.10} &{\bfseries 9.21}\\
 \hline 
\end{tabular}
\end{table}
\FloatBarrier
\begin{table*}[htbp!]
\caption{ablation study of contributions with the averaged speech enhancement performance of three snr levels (-5, 0, and 5 dB), three noise interferences ($factory$, $babble$, and $cafe$), and three room environments (ipad$\char`_$livingroom1, ipad$\char`_$bedroom1, and ipad$\char`_$confroom1). each result is the average value of 5400 experiments.}
\centering
%\small\addtolength{\tabcolsep}{-1pt}
\begin{tabular}{ccccccccc}
\hline
\multicolumn{3}{c}{Ablation Settings} & %
   \multirow{2}{*}{Training Time (h)}& \multirow{2}{*}{PESQ}     & \multirow{2}{*}{CSIG}& \multirow{2}{*}{CBAK}& \multirow{2}{*}{COVL}& \multirow{2}{*}{SDR (dB)}\\
\cline{1-3}
  Feature Combination& Masking Module&Ensemble Pre-Tasks \\
 \hline
% Unprocessed  & 41.5&1.44 &3.04 & \\
\xmark &\xmark &\xmark &10.7 & 1.48&2.28&1.90&1.84 &4.76 \\
\checkmark   &\xmark &\xmark  &10.8 &1.56 & 2.39 & 1.94&1.88 &5.16  \\
\xmark  &  \checkmark &\xmark  &11.9 &1.71 & 2.45 & 2.16&1.97 &5.41 \\
\xmark  &  \checkmark &\checkmark &12.5 &1.77 & 2.48 & 2.17&2.06 &7.02 \\
\checkmark   &\checkmark  &\xmark &12.0 &1.80 & 2.56 & 2.17&2.02 &6.94 \\
\checkmark   &\checkmark   &\checkmark &12.8 &{\bfseries 1.94} &{\bfseries 2.63} & {\bfseries 2.20}& {\bfseries 2.10}& {\bfseries 9.21}\\
 \hline 
\end{tabular}
\end{table*}
% \begin{table*}[htbp!]
% \caption{ablation study of contributions with the averaged speech enhancement performance of three snr levels (-5, 0, and 5 dB), three noise interferences ($factory$, $babble$, and $cafe$), and three room environments (ipad$\char`_$livingroom1, ipad$\char`_$bedroom1, and ipad$\char`_$confroom1). each result is the average value of 5400 experiments.}
% \centering
% %\small\addtolength{\tabcolsep}{-1pt}
% \begin{tabular}{ccccccccc}
% \hline
% \multicolumn{3}{c}{Ablation Settings} & %
%   \multirow{2}{*}{Training Time (s)}& \multirow{2}{*}{PESQ}     & \multirow{2}{*}{CSIG}& \multirow{2}{*}{CBAK}& \multirow{2}{*}{COVL}& \multirow{2}{*}{SDR (dB)}\\
% \cline{1-3}
%   Feature Combination& Masking Module&Ensemble Pre-Tasks \\
%  \hline
% % Unprocessed  & 41.5&1.44 &3.04 & \\
% \xmark &\xmark &\xmark &38434 & 1.48&2.28&1.90&1.84 &4.76 \\
% \checkmark   &\xmark &\xmark  &39029 &1.56 & 2.39 & 1.94&1.88 &5.16  \\
% \xmark  &  \checkmark &\xmark  &42651 &1.71 & 2.45 & 2.16&1.97 &5.41 \\
% \xmark  &  \checkmark &\checkmark &45121 &1.77 & 2.48 & 2.17&2.06 &7.02 \\
% \checkmark   &\checkmark  &\xmark &43215 &1.80 & 2.56 & 2.17&2.02 &6.94 \\
% \checkmark   &\checkmark   &\checkmark &45943 &{\bfseries 1.94} &{\bfseries 2.63} & {\bfseries 2.20}& {\bfseries 2.10}& {\bfseries 9.21}\\
%  \hline 
% \end{tabular}
% \end{table*}
The speech enhancement performance of various routines can be seen from TABLE VI. The routine 1 is the reproduction of the baseline \cite{ssl} which only learns the latent representation as the single pre-task. The masking module is added as the second pre-task in the routine 2 and improves the speech enhancement compared to the routine 1. For example, in terms of PESQ, the speech enhancement has a 13.3$\%$ improvement with the masking module. As for the routine 3, the learned latent representation is used to train the masking module. Consequently, the target speech feature is well preserved in the enhanced features while interference is effectively reduced such that the CAE generalizes better to limited training data. Compared to the routine 2, the speech enhancement performance of the routine 3 is improved. Conversely, the estimation accuracy of speech and mixture latent representations is refined by the loss of the masking module in the routine 4. The speech enhancement performance of the routines 3$\&$4 is closed e.g., reach 2.06 and 2.04 in terms of COVL, respectively. In the proposed method, the routine 5 combines the routines 3$\&$4. The loss of each pre-task e.g., the latent representation and the masking module in the ensemble learning are exploited to train the other pre-task to train the other pre-task and the performance is further improved.

Furthermore, the effectiveness of each contribution is investigated based on the DAPS dataset. The experimental results in terms of four performance measurements and the training time are shown in TABLE VII. It is highlighted that the recorded time consists of both the feature extraction and networks training. Due to the dependency between the masking module and ensemble pre-tasks, the ablation experiments with the ensemble pre-tasks but without the masking module are not performed.

% based on the DAPS dataset. In the baseline, a single spectrogram is obtained from the output layer of the decoders in CAE and MAE to compare to the proposed multi-resolution method. 

Initially, the effectiveness of the feature combination is studied. We conduct two sets of experiments that differ at the features of input speech and mixtures. First, the spectra are fed into the encoder as the baseline \cite{ssl}. Then, the proposed method has an SDR improvement of 8.4$\%$ after the feature combination is extracted from the spectra. The proposed method assigns the weights to each feature of the feature combination to learn the latent representation of the target feature in a balanced way. Consequently, different information, distributed in various features, is extracted to refine the accuracy of the target speech estimation.

Moreover, the experiment is performed by adding the proposed masking module. From TABLE VII, it can be observed that the performance is significantly improved by the DM and ERM estimation among all four measurements. For example, in terms of PESQ, the performance is improved from 1.48 to 1.71, which further confirms that the proposed method with the masking module can boost the enhancement performance. The use of DM can mitigate the adverse effect of acoustic reflections to extract the target speech from the noisy mixture. Then, the ERM is estimated by using the desired speech and the estimated dereverberated mixture, which can further improve the dereverberation. Thus, the proposed ERM can better model the relationship between the clean speech and the estimated dereverberated mixture. As a result, the proposed masking module has a better ability in adapting to unseen speakers and leading to improved performance in highly reverberant scenarios.

% Moreover, the experiment is performed by adding the proposed phase-aware decoders. From TABLE VI, it can be observed that the performance is significantly improved by the proposed phase-aware method among all four measurements. For example, in terms of PESQ, the performance is improved from 1.56 to 1.69, which further confirms that the proposed method with the phase-aware decoders can boost the enhancement performance. In the baselines, the speech signal is reconstructed by using the noisy phase and the estimated magnitude, which causes a phase loss between the clean speech signal and the corresponding reconstruction. However, the proposed phase-aware method utilizes $D_{12}$ and $D_{22}$ to estimate the phase of the target speech signal and speech mixture, respectively, and improve the accuracy of estimation.

The ensemble learning i.e., the routine 5 is introduced to the proposed method. Compared to the baselines, the proposed ensemble learning brings an obvious improvement in terms of all performance measurements. For instance, the proposed method has a PESQ improvement from 1.48 to 1.71 after the ensemble learning is introduced. In the SSL study, due to the limited training data, the learned information from the latent representation and the masking module is shared between the pre-tasks and plays an important role in the speech enhancement problem. With the proposed ensemble learning, each of the pre-task is estimated with the updated reconstruction of the other and the desired speech information is better preserved in the enhanced features.

Furthermore, the training time of models with each contribution is presented in TABLE VII. The computational cost is increased by exploiting contributions to the proposed method. Therefore, there is a trade-off between the computational cost and the speech enhancement performance.
\section{CONCLUSION}
In this paper, we proposed an SSL method with the feature combination and ensemble pre-tasks to solve the monaural speech enhancement problem. We demonstrated that various features showed different performances in the SSL case. The learned information of each feature was assigned with different weights and combined to estimate the target speech and mixture spectra. Then, the masking module was added as the second pre-task and further improved the speech enhancement performance. Moreover, we provided five training routines and selected the routine 5 i.e., shared the learned information between two pre-tasks. The experimental results showed that the proposed method outperformed the state-of-the-art SSL approaches.

To further improve the performance and reduce the computational cost, one direction is to divide the noisy mixture spectra into two subbands and use more computational cost on the lower-band where the signal energy is more than the upper-band \cite{IET}. Besides, the proposed method reconstructs the target speech by using the noisy phase and the estimated magnitude. Future work should be dedicated to estimating both the amplitude and phase of the mixture feature to further refine the speech enhancement performance.

% or


% \appendices

% % you can choose not to have a title for an appendix
% % if you want by leaving the argument blank
% \section{}

% % use section* for acknowledgment
% \section*{Acknowledgment}


\ifCLASSOPTIONcaptionsoff
  \newpage
\fi



\bibliographystyle{IEEEtran}
%\bibliographystyle{IEEEbib}
\bibliography{Yang_Sun}

\begin{thebibliography}{10}
\providecommand{\url}[1]{#1}
\csname url@samestyle\endcsname
\providecommand{\newblock}{\relax}
\providecommand{\bibinfo}[2]{#2}
\providecommand{\BIBentrySTDinterwordspacing}{\spaceskip=0pt\relax}
\providecommand{\BIBentryALTinterwordstretchfactor}{4}
\providecommand{\BIBentryALTinterwordspacing}{\spaceskip=\fontdimen2\font plus
\BIBentryALTinterwordstretchfactor\fontdimen3\font minus
  \fontdimen4\font\relax}
\providecommand{\BIBforeignlanguage}[2]{{%
\expandafter\ifx\csname l@#1\endcsname\relax
\typeout{** WARNING: IEEEtran.bst: No hyphenation pattern has been}%
\typeout{** loaded for the language `#1'. Using the pattern for}%
\typeout{** the default language instead.}%
\else
\language=\csname l@#1\endcsname
\fi
#2}}
\providecommand{\BIBdecl}{\relax}
\BIBdecl

\bibitem{tomo}
R.~Ikeshita, K.~Kinoshita, N.~Kamo, and T.~Nakatani, ``{Online speech
  dereverberation using mixture of multichannel linear prediction models},''
  \emph{IEEE signal processing letters}, vol.~28, pp. 1580 -- 1584, 2021.

\bibitem{TAI}
Y.~Li, Y.~Sun, K.~Horoshenkov, and S.~M. Naqvi, ``{Domain adaptation and
  autoencoder based unsupervised speech enhancement},'' \emph{IEEE Transactions
  on Artificial Intelligence}, vol.~3, no.~1, pp. 43 -- 52, 2021.

\bibitem{ssl}
Y.-C. Wang, S.~Venkataramani, and P.~Smaragdis, ``{Self-supervised learning for
  speech enhancement},'' \emph{International Conference on Machine Learning
  (ICML)}, 2020.

\bibitem{ssl4}
B.~Dendani, H.~Bahi, and T.~Sari, ``{Self-supervised speech enhancement for
  arabic speech recognition in real-world environments},'' \emph{Traitement du
  Signal}, vol.~38, no.~2, pp. 349 -- 358, 2021.

\bibitem{ranlp}
T.~Nayak, N.~Majumder, and S.~Poria, ``Improving distantly supervised relation
  extraction with self-ensemble noise filtering,'' \emph{Recent Advances in
  Natural Language Processing}, 2021.

\bibitem{datap}
A.~Ratner, C.~D. Sa, S.~Wu, D.~Selsam, and C.~Ré, ``{Data programming:
  creating large training sets, quickly},'' \emph{Neural Information Processing
  Systems}, vol.~29, pp. 3567 -- 3575, 2016.

\bibitem{jointi}
Z.~H. Chen, J.~Droppo, J.~Y. Li, and W.~Xiong, ``{Progressive joint modeling in
  unsupervised single-channel overlapped speech recognition},'' \emph{IEEE/ACM
  Transactions on Audio Speech and Language Processing}, vol.~26, pp. 184 --
  196, 2018.

\bibitem{hubert}
W.-N. Hsu, B.~Bolte, Y.-H.~H. Tsai, K.~Lakhotia, R.~Salakhutdinov, and
  A.~Mohamed, ``{HuBERT: self-supervised speech representation learning by
  masked prediction of hidden units},'' \emph{arXiv preprint arXiv:2106.07447},
  2021.

\bibitem{ssl6}
R.~E. Zezario, T.~Hussain, X.~G. Lu, H.-M. Wang, and Y.~Tsao,
  ``{Self-supervised denoising autoencoder with linear regression decoder for
  speech enhancement},'' \emph{IEEE International Conference on Acoustics,
  Speech and Signal Processing (ICASSP)}, 2020.

\bibitem{Two}
Y.~Sun, W.~Wang, J.~A. Chambers, and S.~M. Naqvi, ``Two-stage monaural source
  separation in reverberant room environments using deep neural networks,''
  \emph{IEEE/ACM Transactions on Audio, Speech, and Language Processing},
  vol.~27, no.~1, pp. 125--138, 2019.

\bibitem{IET}
Y.~Li, Y.~Sun, and S.~M. Naqvi, ``{Single-channel dereverberation and denoising
  based on lower band trained SA-LSTMs},'' \emph{IET Signal Processing},
  vol.~14, no.~10, pp. 774 -- 782, 2021.

\bibitem{dsp}
Y.~Sun, L.~Zhu, J.~A. Chambers, and S.~M. Naqvi, ``{Monaural source separation
  based on adaptive discriminative criterion in neural networks},''
  \emph{International Conference on Digital Signal Processing (DSP)}, 2017.

\bibitem{complexirm}
Y.~Zhang, Y.~Liu, and D.~L. Wang, ``{Complex ratio masking for singing voice
  separation},'' \emph{IEEE International Conference on Acoustics, Speech and
  Signal Processing (ICASSP)}, 2021.

\bibitem{featurezuoyong}
W.~Han, C.~M. Wu, X.~W. Zhang, M.~Sun, and G.~Min, ``{Speech enhancement based
  on improved deep neural networks with MMSE pretreatment features},''
  \emph{IEEE International Conference on Signal Processing (ICSP)}, 2016.

\bibitem{featureset}
Y.~X. Wang, K.~Han, and D.~L. Wang, ``{Exploring monaural features for
  classification-based speech segregation},'' \emph{IEEE Transactions on Audio,
  Speech, and Language Processing}, vol.~21, no.~2, pp. 270--279, 2013.

\bibitem{CSA1}
Y.~Sun, Y.~Xian, W.~Wang, and S.~M. Naqvi, ``Monaural source separation in
  complex domain with long short-term memory neural network','' \emph{IEEE
  Journal of Selected Topics in Signal Processing}, vol.~13, no.~2, pp. 359 --
  369, 2019.

\bibitem{mfcc}
M.~Xu, L.-Y. Duan, J.~F. Cai, L.-T. Chia, C.~S. Xu, and Q.~Tian, ``{HMM-based
  audio keyword generation},'' \emph{Advances in Multimedia Information
  Processing: 5th Pacific Rim Conference on Multimedia}, pp. 566 -- 574, 2004.

\bibitem{mfcc1}
R.-W. Li, X.~Y. Sun, T.~Li, and N.~Z. F, ``{A multi-objective learning speech
  enhancement algorithm based on IRM post-processing with joint estimation of
  SCNN and TCNN},'' \emph{Digital Signal Processing}, vol. 101, p. 102731,
  2020.

\bibitem{ams}
B.~Kollmeier and R.~Koch, ``Speech enhancement based on physiological and
  psychoacoustical models of modulation perception and binaural interaction,''
  \emph{The Journal of the Acoustical Society of America}, vol.~95, no.~3, pp.
  1593--1602, 1994.

\bibitem{ams3}
Y.~Wang and M.~Brookes, ``{Speech enhancement using an MMSE spectral amplitude
  estimator based on a modulation domain Kalman filter with a Gamma prior},''
  \emph{IEEE International Conference on Acoustics, Speech and Signal
  Processing (ICASSP)}, 2016.

\bibitem{ams4}
N.~Moritz, J.~Anemüller, and B.~Kollmeier, ``{Amplitude modulation spectrogram
  based features for robust speech recognition in noisy and reverberant
  environments},'' \emph{IEEE International Conference on Acoustics, Speech and
  Signal Processing (ICASSP)}, 2011.

\bibitem{PLP2}
H.~Hermansky and N.~Morgan, ``{RASTA processing of speech},'' \emph{IEEE
  Transactions on Audio, Speech, and Language Processing}, vol.~2, no.~4, pp.
  578--589, 1994.

\bibitem{RASTA}
H.~Hermansky, N.~Morgan, and H.-G. Hirsch, ``{Recognition of speech in additive
  and convolutive noise based on RASTA spectral processing},'' \emph{IEEE
  International Conference on Acoustics, Speech and Signal Processing
  (ICASSP)}, 1993.

\bibitem{RASTA2}
M.~A.~A. Zulkifly and N.~Yahya, ``{Relative spectral-perceptual linear
  prediction (RASTA-PLP) speech signals analysis using singular value
  decomposition (SVD)},'' \emph{IEEE 3rd International Symposium on Robotics
  and Manufacturing Automation (ROMA)}, 2017.

\bibitem{coch}
R.~V. Sharana and T.~J. Moir, ``{Pseudo-color cochleagram image feature and
  sequential feature selection for robust acoustic event recognition},''
  \emph{Applied Acoustics}, vol. 140, pp. 198--204, 2018.

\bibitem{ssl5}
A.Sivaraman and M.~Kim, ``{Self-Supervised learning for personalized speech
  enhancement},'' \emph{arXiv preprint arXiv:2104.02017}, 2020.

\bibitem{ssl7}
Y.-C. Chen, S.-W. Yang, C.-K. Lee, S.~See, and H.-Y. Lee, ``Speech
  representation learning through self-supervised pretraining and multi-task
  finetuning,'' \emph{arXiv preprint arXiv:2110.09930}, 2021.

\bibitem{cl}
A.~Sivaraman and M.~Kim, ``Self-supervised learning from contrastive mixtures
  for personalized speech enhancement,'' \emph{arXiv preprint
  arXiv:2011.03426}, 2020.

\bibitem{icmla}
T.~Sun, S.~Y. Gong, Z.~W. Wang, C.~D. Smith, X.~H. Wang, L.~Xu, and J.~D. Liu,
  ``{Boosting the intelligibility of waveform speech enhancement networks
  through self-supervised representations},'' \emph{IEEE International
  Conference on Machine Learning and Applications (ICMLA)}, 2021.

\bibitem{shouer}
H.-S. Choi, J.~Lee, W.~Kim, J.~H. Lee, H.~Heo, and K.~Lee, ``{Neural analysis
  and synthesis: reconstructing speech from self-supervised representations},''
  \emph{Neural Information Processing Systems (NeurIPS)}, 2021.

\bibitem{Student}
D.~L. Wang and J.~T. Chen, ``{Supervised speech separation based on deep
  learning: An overview},'' \emph{IEEE/ACM Transactions on Audio, Speech, and
  Language Processing}, vol.~26, pp. 1702--1726, 2018.

\bibitem{features}
Y.~X. Wang, K.~Han, and D.~L. Wang, ``{Exploring monaural features for
  classification-based speech segregation},'' \emph{IEEE Transactions on Audio,
  Speech, and Language Processing}, vol.~21, no.~2, pp. 270--279, 2013.

\bibitem{delta}
A.~V. Haridas, R.~Marimuthu, and B.~Chakraborty, ``A novel approach to improve
  the speech intelligibility using fractional delta-amplitude modulation
  spectrogram,'' \emph{Cybernetics and Systems}, vol.~49, no. 7-8, pp.
  421--451, 2018.

\bibitem{filter}
J.~T. Chen, Y.~X. Wang, and D.~L. Wang, ``{A feature study for classification
  based speech separation at very low signal-to-noise ratio},'' \emph{IEEE
  International Conference on Acoustics, Speech and Signal Processing
  (ICASSP)}, 2014.

\bibitem{receptive}
M.~Hasannezhad, Z.~H. Ouyang, W.-P. Zhu, and B.~Champagne, ``{An integrated
  CNN-GRU framework for complex ratio mask estimation in speech enhancement},''
  \emph{Asia-Pacific Signal and Information Processing Association Annual
  Summit and Conference (APSIPA ASC)}, 2020.

\bibitem{bn}
S.~Ioffe and C.~Szegedy, ``{Batch normalization: accelerating deep network
  training by reducing internal covariate shift},'' \emph{International
  Conference on Machine Learning (ICML)}, 2015.

\bibitem{ssl1}
S.-F. Huang, S.-P. Chuang, D.-R. Liu, Y.-C. Chen, G.-P. Yang, and H.-Y. Lee,
  ``{Stabilizing label assignment for speech separation by self-supervised
  pre-training},'' \emph{Interspeech}, 2021.

\bibitem{mama}
A.~T. Liu, S.-W. Li, and H.-Y. Lee, ``{TERA: self-supervised learning of
  transformer encoder representation for speech},'' \emph{IEEE/ACM Transactions
  on Audio, Speech, and Language Processing}, vol.~29, pp. 2351 -- 2366, 2021.

\bibitem{wav}
A.~Baevski, H.~Zhou, A.~Mohamed, and M.~Auli, ``{wav2vec 2.0: a framework for
  self-supervised learning of speech representations},'' \emph{Neural
  Information Processing Systems (NeurIPS)}, 2020.

\bibitem{DT}
J.~J. Chen, Q.~R. Mao, and D.~Liu, ``{Dual-path transformer network: direct
  context-aware modeling for end-to-end monaural speech separation},''
  \emph{Interspeech}, 2020.

\bibitem{daps}
G.~J. Mysore, ``{Can we automatically transform speech recorded on common
  consumer devices in real-world environments into professional production
  quality speech?—a dataset, insights, and challenges},'' \emph{IEEE Signal
  Processing Letters}, vol.~22, no.~8, pp. 1006 -- 1010, 2014.

\bibitem{noise}
A.~Varga and H.~J.~M. Steeneken, ``Assessment for automatic speech recognition:
  Ii. noisex-92: a database and an experiment to study the effect of additive
  noise on speech recognition systems,'' \emph{IEEE Transactions on Audio,
  Speech, and Language Processing}, vol.~12, no.~3, pp. 247 -- 251, 1993.

\bibitem{mos}
Y.~Hu and P.~C. Loizou, ``{Evaluation of objective quality measures for
  speech},'' \emph{IEEE Transactions on Audio, Speech, and Language
  Processing}, vol.~16, no.~1, pp. 229 -- 238, 2008.

\bibitem{lib}
J.~Cosentino, M.~Pariente, S.~Cornell, A.~Deleforge, and E.~Vincent,
  ``{LibriMix: an open-source dataset for generalizable speech separation},''
  \emph{Interspeech}, 2020.

\bibitem{AMS2}
G.~Kim, Y.~Lu, Y.~Hu, and P.~C. Loizou, ``{An algorithm that improves speech
  intelligibility in noise for normal-hearing listeners},'' \emph{Journal of
  the Acoustical Society of America}, vol. 126, pp. 1486--1494, 2009.

\bibitem{mapping}
Y.~Xu, J.~Du, L.-R. Dai, and C.-H. Lee, ``A regression approach to speech
  enhancement based on deep neural networks,'' \emph{IEEE/ACM Transanctions on
  Audio Speech and Language Processing}, vol.~23, no.~1, pp. 7--19, 2015.

\bibitem{IRM1}
Y.~Wang, A.~Narayanan, and D.~L. Wang, ``{On training targets for supervised
  speech separation},'' \emph{IEEE/ACM Transactions on Audio, Speech, and
  Language Processing}, vol.~22, no.~12, pp. 1849--1858, 2014.

\bibitem{cirm}
D.~S. Williamson and D.~L. Wang, ``Time-frequency masking in the complex domain
  for speech dereverberation and denoising,'' \emph{IEEE/ACM Transanctions on
  Audio Speech and Language Processing}, vol.~25, no.~7, pp. 1492--1501, 2017.

\end{thebibliography}
\end{document}